\documentclass{article}
\usepackage{hyperref}
\usepackage[utf8]{inputenc}
\usepackage{latexsym,graphicx,amssymb,amsmath}
\usepackage{cite}
\usepackage{float}

\newcommand{\bea}{\begin{eqnarray}}
\newcommand{\eea}{\end{eqnarray}}
\newcommand{\be}{\begin{equation}}
\newcommand{\ee}{\end{equation}}

\begin{document}

\title{\textbf {On positronium states in ${\rm{QED_3}}$ }}

\author{Chang-Yong Liu\footnote{email address:
liuchangyong@nwsuaf.edu.cn},\ \ You-Wei Li\footnote{email address: 16317163@qq.com}\ \ and\ \ Su-Zhi Wu \\
College of Science, Northwest A\&F University, Yangling, Shaanxi
712100, China\\
}

\maketitle

\begin{abstract}
In this note, we present a new method to investigate the positronium states in ${\rm{QED_3}}$. According to the $\rm{K\ddot{a}ll\acute{e}n-Lehmann}$ spectral representation, the energy eigenvalues of bound states are poles of the correlation function. Using the chain approximation, we obtain the energy eigenvalues of the vector positronium states by taking into account the single-valued branches of multi-valued function. Using the same method, we also find the electron physical mass at some single-valued branch of multi-value function.
Our results are agreement with the known ones.
\end{abstract}

\section{Introduction}
Quantum electrodynamics in 2+1 dimensions (${\rm{QED_3}}$) is an interesting gauge field theory. The theory is super-renormalizable and connected to quantum chromodynamics (QCD) in 3+1 dimensions \cite{Cornwall:1980zw}.  One of the most interesting features of ${\rm{QED_3}}$ is that the photon can have a topological mass term called a Chern-Simons term \cite{Jackiw:1980kv,Schonfeld:1980kb,Deser:1981wh}. ${\rm{QED_3}}$ is an abelian theory and has a confining logarithmic potential \cite{Burden:1991uh}.  In 2+1 dimensions, the potential of the $e^+e^-$ due to one-photon exchange need
a regulating photon mass $\mu$ \cite{Cornwall:1980zw,Sen:1990jm}
\bea
V(r)=-e^2\int \frac{d^2k}{(2\pi)^2}\frac{e^{i\mathbf{k}\cdot \mathbf{r}}}{k^2+\mu^2}=-\frac{e^2}{2\pi}K_0
(|\mu| r).\nonumber
\eea
In the limit $\mu \rightarrow 0 $, the potential becomes
\bea
V(r)=\frac{e^2}{2\pi}\ln(\frac{\mu r}{2})+O(1).\nonumber
\eea
 On the other hand, the renormalized  mass $m_R$ in one-loop is
\bea
m_R=m+\frac{e^2}{4\pi} \ln(\frac{m}{\mu}).\nonumber
\eea
Where the $m$ is the bare electron mass.
Then the $V(r)$ and $m_R$ are infrared divergent. But the infrared divergences cancel in the sum of
$2m_R$ and $V(r)$
\bea
2m_R+V(r)=2m+\frac{e^2}{2\pi}\ln(\frac{m r}{2})+O(1).\nonumber
\eea
To study the positronium states in ${\rm{QED_3}}$, we need to solve the Schr\"{o}dinger equation with this potential. The non-relativistic Coulomb Schr\"{o}dinger equation for ${\rm{QED_3}}$ with the confining logarithmic potential is derived from the LCQ formalism \cite{Yung:1991tz,Tam:1994qk}. The other approach to positronium states is via a solution to the homogenous Bethe-Salpeter equation \cite{Salpeter:1951sz} with fermion propagator input from the Schwinger-Dyson equation \cite{Allen:1995ui,Allen:1996ri}. Their non-relativistic position-space result for the positronium states is
\bea
[-\frac{1}{m}\nabla^2+\frac{1}{2\pi}(C+\ln(mr))]\psi(\vec{r})=(E-2m)\psi(\vec{r}),\nonumber
\eea
where $C$ is Euler's constant. The expression for the bound state energy is \cite{Tam:1994qk,Koures:1995qp}
\bea \label{em2}
E_n^l=2m+\frac{1}{4\pi}\ln m+\frac{1}{2\pi}(\lambda_n^l-\frac{1}{2}\ln \frac{2}{\pi}).
\eea
Where the $l$ is the orbital angular momentum. There are first three eigenvalues for $l$ ranging from 0 to 2  in Table \ref{five}.
\begin{table}[htbp]
  \centering
  \begin{tabular}{|c|c|c|c|c|c|}
\hline
  $$&$\lambda_0^l$&$\lambda_1^l$&$\lambda_2^l$&$\lambda_3^l$&$\lambda_4^l$  \\
\hline
$l=0$ &1.7969 &2.9316 &3.4475 &3.7858 &4.0380  \\
  \hline
  $l=1$ &2.6566&3.2798&3.6647&3.9430&4.1610 \\
  \hline
  $l=2$             &3.1147&3.5462&3.8504&4.0848&4.2753 \\
\hline
  \end{tabular}
  \caption{\label{five} First three eigenvalues for $l$ ranging from 0 to 2 \cite{Tam:1994qk,Koures:1995qp}. }
  \end{table}

\begin{figure}[H]
\center
\includegraphics[width=0.2\textwidth]{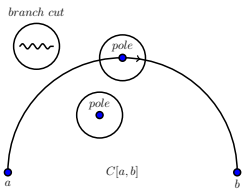}
\caption{\label{2} A smooth contour $C[a,b]$ in complex plane staring from $a$ to $b$. The blue dots and wave line denote the poles and branch cut  of function $f(z)$ separately.  }
\end{figure}

Following our previous work \cite{Liu:2018dfm}, we use a new method to study the positronium states in ${\rm{QED_3}}$. Our approach is using the analytical structure of the correlation function. The exact Feynman propagator for the gauge field in the $\rm{K\ddot{a}ll\acute{e}n-Lehmann}$ spectral representation \cite{Kallen:1952zz} is given by
\be \label{KL}
\langle\Omega| T(A_{\mu}(x)A_{\nu}(y))|\Omega\rangle=\frac{1}{2}\int_0^{\infty} dm^2 \rho_{\mu\nu}(m^2)\Delta_F(x-y;m^2).
\ee
The pole of the Fourier transform of equation (\ref{KL}) gives the mass of the bound state.
In order to study the bound states, we define the integral of a complex function $f(z)$ along a smooth contour $C[a,b]$ in complex plane. Suppose the function $f(z)$ have poles or branch cuts (FIG. \ref{2}), then the integral of $f(z)$ along the contour $C[a,b]$ can be expressed as
\par

\bea
\label{fz}
\int_{C[a,b]}f(z)dz=P\int_a^bf(z)dz+\sum n_i \oint_{C_i}f(z)dz.
\eea

Where the $C_i$ is a closed curve circling the pole or branch cut. The $P\int_a^bf(z)dz$ takes value in main single-valued branch. The winding number $n_i\in Z$ is the
contour circling $n_i$ times around the pole or branch cut. In our previous work \cite{Liu:2020dfm}, we recalculated the axial (ABJ) anomaly \cite{Adler:1969gk,Bell:1969ts} by the formulae (\ref{fz}). We found the divergence of the
axial current which is
\bea \label{bnc}
q^{\rho}M_5^{\rho\mu\nu}&=&\frac{m^2}{2\pi^2}\epsilon^{\mu\nu}_{\quad \rho\sigma}k_1^{\rho}k_2^{\sigma}[\frac{1}{q^2}P\int_0^1 \frac{1}{x}\ln \frac{1}{1-x(1-x)\frac{q^2}{m^2}}dx+\frac{1}{q^2}2\pi i k
\ln\frac{1+\sqrt{1-\frac{4m^2}{q^2}}}{1-\sqrt{1-\frac{4m^2}{q^2}}}+\frac{1}{q^2}(2\pi i)^2 n] \nonumber\\
&-&
\frac{1}{4\pi^2}\epsilon^{\mu\nu}_{\quad \rho\sigma}k_1^{\rho}k_2^{\sigma}+\left(
  \begin{array}{c}
    \mu\leftrightarrow\nu \\
    k_1\leftrightarrow k_2 \\
  \end{array}
\right).
\eea
Where the $n$ and $k$ are $n\in Z$ and $k\in Z$. We
 used the extra term $\frac{m^2}{2\pi^2}\epsilon^{\mu\nu}_{\quad \rho\sigma}k_1^{\rho}k_2^{\sigma}[-\frac{4\pi^2}{q^2}n]$ in equation (\ref{bnc}) to cancel
 the anomaly term $-
\frac{1}{4\pi^2}\epsilon^{\mu\nu}_{\quad \rho\sigma}k_1^{\rho}k_2^{\sigma}$. This leaded to the anomaly free condition
\be
\label{boundstates}
\boldsymbol{q^2=8n\pi^2m^2=m_P^2(n)}, \quad n\in N.
\ee
Where $m_P$ is the neutral pseudoscalar meson mass for a quantum number $n$. The
single-valued branches of multi-valued function are related to the bound states.

The paper is organized as follows. In Section 2, we study the positronium ($e^+e^-$) systems in ${\rm{QED_3}}$ with two-component Dirac fermion. We end with the conclusions. \\

\section{Positronium ($e^+e^-$) systems in ${\rm{QED_3}}$ with two-component Dirac fermion}
\begin{figure}[H]
\center
\includegraphics[width=0.2\textwidth]{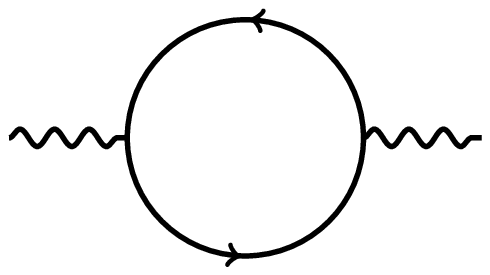}
\caption{\label{oneloop} The photon propagator with a single fermion loop insertion.}
\end{figure}

In this section, we consider the ${\rm{QED_3}}$ with single two-component Dirac fermion. The Lagrangian density of the theory is given by
\be
\mathcal{L}=-\frac{1}{4}F_{\mu\nu}F^{\mu\nu}+\bar{\psi}(i\partial\!\!\!/-eA\!\!\!/-m)\psi.\nonumber
\ee
We use the Minkowski metric tensor $g^{\mu\nu}={\rm{diag(1,-1,-1)}}$. The Dirac gamma matrices are defined by $\gamma^0=\sigma_3, \gamma^1=i\sigma_1, \gamma^2=i\sigma_2$, where the
$\sigma_i$'s are the Pauli matrices. The $2\times2$ Dirac matrices satisfy the identities:
\bea
\gamma^{\mu}\gamma^{\nu}=g^{\mu\nu}\textbf{1}-i\epsilon^{\mu\nu\rho}\gamma_{\rho},\\
{\rm{tr}}(\gamma^{\mu}\gamma^{\nu}\gamma^{\rho})=-2i\epsilon^{\mu\nu\rho},\nonumber
\eea
where we define the totally antisymmetric tensor $\epsilon^{\mu\nu\rho}$ so that $\epsilon^{012}=1$. Different with the 3 + 1 dimensional theories, the trace of three gamma matrices in 2+1 dimensions produces the totally antisymmetric
$\epsilon^{\mu\nu\rho}$ symbol. We define $i\Pi^{\mu\nu}(q)$ to be the sum of all 1-particle-irreducible (1PI) insertions into the photon propagator. The expression of $i\Pi_{\mu\nu}(q)$  for one-loop amplitude (Fig. \ref{oneloop}) is
\par

\bea
i\Pi_{\mu\nu}(q)=(-ie)^2(-1)\int\frac{d^3k}{(2\pi)^3}{\rm{tr}}[\gamma_{\mu}\frac{i}{k\!\!\!/-m}\gamma_{\nu}\frac{i}{k\!\!\!/+q\!\!\!/-m}].\nonumber
\eea

Using the same method as ${\rm{QED_4}}$\cite{Pauli:1949zm,Schwinger:1951nm,Peskin:1995ev}, we obtain

\bea \label{pi}
\Pi_{\mu\nu}(q)=(g_{\mu\nu}-\frac{q_{\mu}q_{\nu}}{q^2})\Pi_1(q^2)+im\epsilon_{\mu\nu\rho}q^{\rho}\Pi_2(q^2),
\eea
where the expression of $\Pi_1(q^2)$ and $\Pi_1(q^2)$ are
\bea
\Pi_1(q^2)&=&4ie^2q^2\int_0^1dx\ x(1-x)\int\frac{d^3k}{(2\pi)^3}\frac{1}{(k^2-\Delta)^2},\nonumber\\
\Pi_2(q^2)&=&-2ie^2\int_0^1dx\int\frac{d^3k}{(2\pi)^3}\frac{1}{(k^2-\Delta)^2}.\nonumber
\eea
The $\Delta$ is defined as
\bea
\Delta=m^2-x(1-x)q^2.\nonumber
\eea
We Wick-rotate and substitute the Euclidean variable $k^0_E=-ik^0$. This gives
\bea
\Pi_1(q^2)&=&-4e^2q^2\int_0^1dx\ x(1-x)\int\frac{d\Omega_3}{(2\pi)^3}\int_0^{\infty} dk_E\frac{k^2_E }{(k_E^2+\Delta)^2}\nonumber\\
&=&-4e^2q^2\int_0^1dx\ x(1-x)\int\frac{d\Omega_3}{(2\pi)^3}\int_0^{\infty} dk_E [\frac{1}{2i\sqrt{\Delta}}(\frac{1}{k_E-i\sqrt{\Delta}}-\frac{1}{k_E+i\sqrt{\Delta}})
-\frac{\Delta }{(k_E^2+\Delta)^2}].\nonumber
\eea
There are two poles at $k_E=\pm i\sqrt{\Delta}$. According to the formulae (\ref{fz}), we obtain
\bea \label{pi1}
\Pi_1(q^2)=\frac{e^2m}{16\pi t}[4t+(4+t^2)\log(\frac{2-t}{2+t})](1+4n),
\eea
where the $n$ is $n\in Z$ and $t^2$ is defined as $t^2=\frac{q^2}{m^2}$. The ${\rm{QED_3}}$ is ultraviolet finite \cite{Jackiw:1980kv,DelCima:2013gpa,DelCima:2015oma}, we don't need a soliton contribution to calculate the positronium state (different with the ultraviolet divergent theories \cite{Liu:2018dfm}). The $\Pi_2(q^2)$ can be calculated with the same method, that is
\bea \label{pi2}
\Pi_2(q^2)&=&-2ie^2\int_0^1dx\int\frac{d^3k}{(2\pi)^3}\frac{1}{(k^2-\Delta)^2}\nonumber\\
&=&\frac{e^2}{4\pi mt}\log(\frac{2+t}{2-t})(1+4k),
\eea
where the $k$ is $k\in Z$.

\begin{figure}[H]
\center
\includegraphics[width=0.6\textwidth]{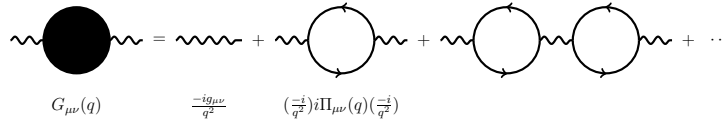}
\caption{\label{vacuum} The photon propagator by the chain approximation.}
\end{figure}

With the chain approximation (Figure \ref{vacuum}), the photon propagator $G_{\mu\nu}(q)$ is then given by
\bea
G_{\mu\nu}(q)&=&\frac{-ig_{\mu\nu}}{q^2}+(\frac{-i}{q^2})i\Pi_{\mu\nu}(q)(\frac{-i}{q^2})+(\frac{-i}{q^2})i\Pi_{\mu}^{\eta}(q)(\frac{-i}{q^2})i\Pi_{\eta\nu}(q)(\frac{-i}{q^2})
+\ldots\nonumber\\
&=&\frac{-ig_{\mu\nu}}{q^2}+\frac{1}{q^2}\Pi_{\mu\eta}(q)G_{\nu}^{\eta}(q).\nonumber
\eea
The dots indicate the iteration of the vacuum polarization tensor.

\par

From this expression, we find the $G_{\mu\nu}(q)$ satisfy the equation
\bea
(q^2g_{\mu\eta}-\Pi_{\mu\eta}(q))G_{\nu}^{\eta}(q)=-ig_{\mu\nu}.\nonumber
\eea

\begin{figure}[H]
\center
\includegraphics[width=0.3\textwidth]{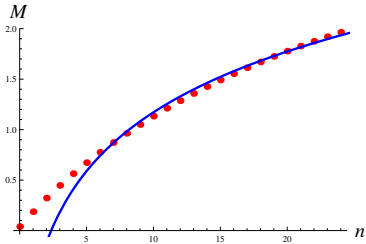}
\caption{\label{QEDtn} Bound state masses $M(0.5,1,n)$. Blue curve is the function $0.9\log[0.35(n+\frac{1}{2})]$.}
\end{figure}

\begin{figure}[H]
\center
\includegraphics[width=0.3\textwidth]{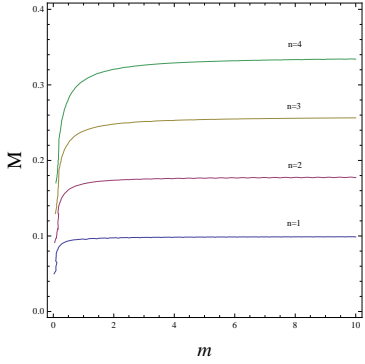}
\caption{\label{QEDem} Bound state masses $M(0.25,m,n)$.  }
\end{figure}

\begin{figure}[H]
\center
\includegraphics[width=0.3\textwidth]{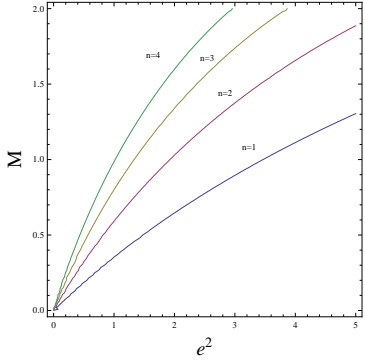}
\caption{\label{QEDme} Bound state masses $M(e^2,1,n)$. }
\end{figure}

After replacing $\Pi_{\mu\eta}(q)$ with the expression (\ref{pi}), we obtain
\bea \label{eq1}
[q^2g_{\mu\eta}-(g_{\mu\eta}-\frac{q_{\mu}q_{\eta}}{q^2})\Pi_1(q^2)-im\epsilon_{\mu\eta\rho}q^{\rho}\Pi_2(q^2)]G_{\nu}^{\eta}(q)=-ig_{\mu\nu}.
\eea
To solve the $G_{\mu\nu}(q)$, we suppose that the $G_{\mu\nu}(q)$ is
\bea
G_{\mu\nu}(q)=g_{\mu\nu}\widetilde{\Pi}_1(q^2)-\frac{q_{\mu}q_{\nu}}{q^2}\widetilde{\Pi}_2(q^2)+im\epsilon_{\mu\nu\rho}q^{\rho}\widetilde{\Pi}_3(q^2),\nonumber
\eea
where the $\widetilde{\Pi}_1(q^2)$, $\widetilde{\Pi}_2(q^2)$ and $\widetilde{\Pi}_3(q^2)$ are unknown functions.
Then the equation (\ref{eq1}) becomes
\bea \label{eeee}
g_{\mu\nu}[q^2-\Pi_1(q^2)]\widetilde{\Pi}_1(q^2)-q_{\mu}q_{\nu}[\widetilde{\Pi}_2(q^2)&-&\frac{\Pi_1(q^2)\widetilde{\Pi}_1(q^2)}{q^2}]
-im\epsilon_{\mu\nu\rho}q^{\rho}[\Pi_2(q^2)\widetilde{\Pi}_1(q^2)-q^2\widetilde{\Pi}_3(q^2)
+\Pi_1(q^2)\widetilde{\Pi}_3(q^2)]\nonumber\\
&+&m^2\epsilon_{\mu\eta\rho}q^{\rho}\epsilon^{\eta}_{\nu\lambda}q^{\lambda}\Pi_2(q^2)\widetilde{\Pi}_3(q^2)=-ig_{\mu\nu}.
\eea
Where the $\epsilon_{\mu\eta\rho}q^{\rho}\epsilon^{\eta}_{\nu\lambda}q^{\lambda}$ can be calculated as
\bea
\epsilon_{\mu\eta\rho}q^{\rho}\epsilon^{\eta}_{\nu\lambda}q^{\lambda}=-g_{\mu\nu}q^2+q_{\mu}q_{\nu}.
\eea
From the equation (\ref{eeee}), we obtain the $\widetilde{\Pi}_1(q^2)$, $\widetilde{\Pi}_2(q^2)$ and $\widetilde{\Pi}_3(q^2)$ satisfy the equations
$$
\begin{cases}
[q^2-\Pi_1(q^2)]\widetilde{\Pi}_1(q^2)-m^2q^2\Pi_2(q^2)\widetilde{\Pi}_3(q^2)=-i,\\
\widetilde{\Pi}_2(q^2)-\frac{\Pi_1(q^2)\widetilde{\Pi}_1(q^2)}{q^2}-m^2\Pi_2(q^2)\widetilde{\Pi}_3(q^2)=0,\\
\Pi_2(q^2)\widetilde{\Pi}_1(q^2)-q^2\widetilde{\Pi}_3(q^2)+\Pi_1(q^2)\widetilde{\Pi}_3(q^2)=0.
\end{cases}$$
The $\widetilde{\Pi}_1(q^2)$, $\widetilde{\Pi}_2(q^2)$ and $\widetilde{\Pi}_3(q^2)$ can be solved as
$$
\begin{cases}
\widetilde{\Pi}_1(q^2)=\frac{-i[q^2-\Pi_1(q^2)]}{[q^2-\Pi_1(q^2)]^2-m^2q^2[\Pi_2(q^2)]^2},\\
\widetilde{\Pi}_3(q^2)=\frac{-i\Pi_2(q^2)}{[q^2-\Pi_1(q^2)]^2-m^2q^2[\Pi_2(q^2)]^2},\\
\widetilde{\Pi}_2(q^2)=\frac{\Pi_1(q^2)\widetilde{\Pi}_1(q^2)}{q^2}+m^2\Pi_2(q^2)\widetilde{\Pi}_3(q^2).
\end{cases}$$
Then the pole of photon propagator $G_{\mu\nu}(q)$ is
\bea \label{eeeee}
[q^2-\Pi_1(q^2)]^2-m^2q^2[\Pi_2(q^2)]^2=0.
\eea
The energy eigenvalues of the bound
states are the solutions of the equation (\ref{eeeee}). Using the expression of $\Pi_1(q^2)$ (\ref{pi1}) and $\Pi_2(q^2)$ (\ref{pi2}), the equation (\ref{eeeee}) can be rewritten as
\bea \label{eeeeere}
\boxed{[m^2t^2-\frac{e^2m}{16\pi t}(4t+(4+t^2)\log(\frac{2-t}{2+t}))(1+4n)]^2-[\frac{e^2m}{4\pi}\log(\frac{2+t}{2-t})(1+4n)]^2=0}.
\eea
 Where the $m$ is the bare fermion mass. For simplify our discussion, we omit the unit of $m$ and $e^2$, where the $m$ and $e^2$ have
the dimensions of ${\rm{(mass)}}$. Taking $e^2=0.5$ and $m=1$ for example, the solution of the bound state mass $M(e^2,m,n)$, where $M(e^2,m,n)$ is $M(e^2,m,n)=\sqrt{q^2}=\sqrt{t^2m^2}$, can be obtained (Table \ref{mesonmass} and Figure \ref{QEDtn}). The $M(0.5,1,n)$ have the behaviour $M(0.5,1,n)\sim a\log[b(n+\frac{1}{2})]$ which is the same as the WKB approximation results \cite{Dobroliubov:1993jr}. We also present
the Figures of $M(0.25,m,n)$ and $M(e^2,1,n)$ in Figure \ref{QEDem} and Figure \ref{QEDme} separately.

\begin{table}[htbp]
  \centering
  \begin{tabular}{|c|c|c|c|c|c|c|c|}
\hline
  $n$ &0&1&2&3&4&5&6  \\
\hline
$M$ &0.039&0.187 &0.322 &0.447 &0.564 &0.673 &0.775  \\
  \hline
  $n$ &7&8&9&10&11&12&13 \\
  \hline
  $M$        &0.872&0.964&1.050&1.133&1.212&1.286&1.358\\
\hline
  \end{tabular}
  \caption{\label{mesonmass}Bound state masses $M(0.5,1,n)$ }
  \end{table}

\par

\par
\par
\begin{figure}[H]
\center
\includegraphics[width=0.5\textwidth]{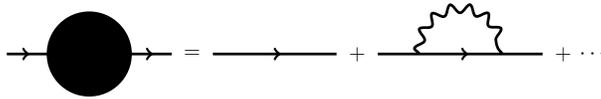}
\caption{\label{fermion2} The full electron propagator.}
\end{figure}

To illustrate our results, we study the physical mass $m_{ph}$ of electron. The electron two-point function can be written as (Figure. \ref{fermion2})

\par
\begin{figure}[H]
\center
\includegraphics[width=0.2\textwidth]{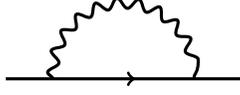}
\caption{\label{fermion} the electron self-energy .}
\end{figure}

\bea
iG(p\!\!\!/)&=&\frac{i}{p\!\!\!/-m}+\frac{i}{p\!\!\!/-m}(i\Sigma(p\!\!\!/))\frac{i}{p\!\!\!/-m}+\cdots \nonumber\\
&=&\frac{i}{p\!\!\!/-m+\Sigma(p\!\!\!/)}.\nonumber
\eea
The $i\Sigma(p\!\!\!/)$ denote the sum of all one-particle irreducible (1PI) diagrams. The leading order term $i\Sigma_2(p\!\!\!/)$ is (Figure. \ref{fermion})

\par
\begin{figure}[H]
\center
\includegraphics[width=0.5\textwidth]{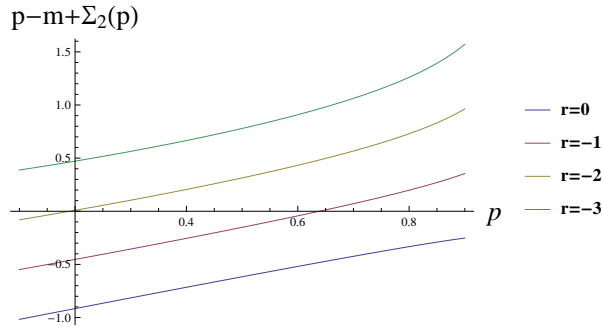}
\caption{\label{cccc} The curve of the function $p-m+\Sigma_2(p)$ with $m=1$, $e^2=0.5$ and $r$ ranging from $-3$ to $0$.}
\end{figure}

\bea
i\Sigma_2(p\!\!\!/)&=&(-ie)^2\int\frac{d^3k}{(2\pi)^3}\ \gamma^{\mu}\frac{i(k\!\!\!/+m)}{k^2-m^2+i\varepsilon}
\gamma_{\mu}\frac{-i}{(k-p)^2+i\varepsilon} \nonumber \\
&=&e^2\int_0^1dx \int \frac{d^3k}{(2\pi)^3}\frac{xp\!\!\!/-3m}{(k^2-\widetilde{\Delta}+i\varepsilon)^2}. \nonumber
\eea
Where the $\widetilde{\Delta}$ is defined as $\widetilde{\Delta}=(1-x)(m^2-p^2x)$. According to the formulae (\ref{fz}), we obtain
\bea
i\Sigma_2(p\!\!\!/)=i\frac{e^2}{8\pi}\int_0^1dx\ \frac{xp\!\!\!/-3m}{\sqrt{\widetilde{\Delta}}}(1+4r),
\eea

 where the $r$ is $r\in Z$.
Then the physical mass $m_{ph}(e^2,m,r)$ is the solution of the equation

\bea \label{edadd}
\boxed {m_{ph}-m+\Sigma_2(m_{ph})=0}.
\eea
We should emphasize that the $i\Sigma_2(p\!\!\!/)$ is free of infrared divergence in our choice of gauge.  From the equation (\ref{edadd}), a real number solution $m_{ph}(e^2,m,r)\in (0,m)$ for some $r\in Z$ exists. Suppose $e^2=0.5$ and $m=1$, the solution of equation (\ref{edadd}) is (Figure \ref{cccc})

$$m_{ph}\approx
\begin{cases}
0.64 \quad r=-1,\\
0.19  \quad r=-2.
\end{cases}$$

\par
We find that the bound state masses $M(e^2,m,n)$ have the excited states $M(e^2,m,n)> 2 m_{ph}$  (Figure \ref{QEDtn} or Table \ref{mesonmass}). This indicate that the
${\rm{QED_3}}$ have properties of confinement.
\par
\begin{figure}[H]
\center
\includegraphics[width=0.5\textwidth]{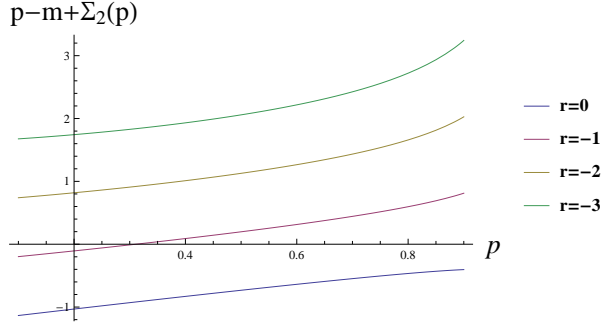}
\caption{\label{cccc1} The curve of the function $p-m+\Sigma_2(p)$ with $m=1$, $e^2=1$ and $r$ ranging from $-3$ to $0$.}
\end{figure}
We now compare our results with the papers\cite{Tam:1994qk,Koures:1995qp} (Table \ref{five}). letting $e^2=1$ and $m=1$, the solution of the physical mass is $m_{ph}(1,1,-1)\approx0.31082$ (Figure \ref{cccc1}). We point out that the $m$ in equation (\ref{em2}) is the physical mass $m_{ph}$ instead of the bare fermion mass, that is
\bea \label{em21}
E_n^l=2m_{ph}+\frac{1}{4\pi}\ln m_{ph}+\frac{1}{2\pi}(\lambda_n^l-\frac{1}{2}\ln \frac{2}{\pi}).
\eea

\begin{figure}[H]
\center
\includegraphics[width=0.5\textwidth]{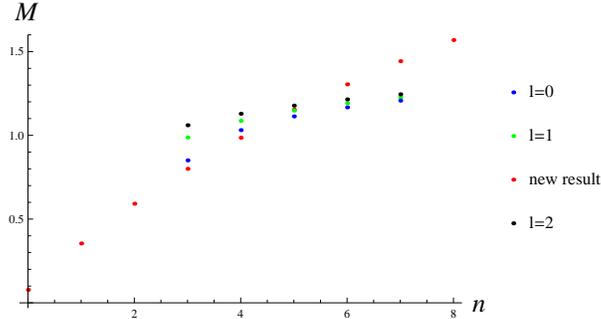}
\caption{\label{cc} The red dots are our results. In here, we let $m=1$ and $e^2=1$. The other dots denote the results in Table 1. The total orbital angular momentum $l$ ranging from 0 to 2.}
\end{figure}

 The solution of the equation (\ref{eeeeere}) have the properties that $M(1,1,n)> 2 m_{ph}(1,1,-1)$ at $n\geq 3$, then $M(1,1,n+3)$ correspond to $E_n^l$ in equation (\ref{em21}) with $n=0,1,2,3$. The results are put in Figure \ref{cc}.
\par
The vector positronium states correspond to the total angular momentum $J=1$ which have related to the total orbital angular momentum $L$ and the total spin $S$ as following
$$J=1 \Leftrightarrow
\begin{cases}
L=0, S=1\\
L=1, S=0 , 1 \\
L=2,S=1
\end{cases}$$

From the Figure \ref{cc}, we see that the vector positronium states which are related to the virtual photons have the total orbital angular momentum $L=0$. Then we put our results and the orbital angular momentum $l=0$ solutions of equation (\ref{em21}) in Table \ref{five11}. From this we find that the first three values in our method are agreement with the ones in \cite{Tam:1994qk,Koures:1995qp}. To make the other two values consistent with each other, we need to calculate the higher order loops of $i\Pi^{\mu\nu}(q)$.

\begin{table}[htbp]
  \centering
  \begin{tabular}{|c|c|c|c|c|c|}
\hline
  $n$&$0$&$1$&$2$&$3$&$4$  \\
\hline
$M(1,1,n+3)$ &0.8000 &0.9858 &1.1530&1.3047 &1.4428  \\
  \hline
  $E^{l=0}_n$ &0.8506&1.0312&1.1133&1.1671&1.2072 \\

\hline
  \end{tabular}
  \caption{\label{five11} Comparing $M(1,1,n+3)$ with $E^{l=0}_n$.  }
  \end{table}


\section{Conclusions and Discussions }
In this note, we have studied the positronium states in ${\rm{QED_3}}$. The results in the calculation of Feynman diagrams are the multi-value functions. Similar to our previous work \cite{Liu:2018dfm,Liu:2020dfm}, these single-valued branches of multi-valued function are related to the bound states. we have calculated the photon propagator by the chain approximation and obtained the equation of energy eigenvalues of the vector positronium states. To illustrate the results, we also studied the electron physical mass in ${\rm{QED_3}}$. The real value electron physical mass $m_{ph}$ have been obtained by considering different single-valued branch of the multi-value function. Our results are agreement with the known ones in \cite{Tam:1994qk,Koures:1995qp}.


\section*{Acknowledgments}
This work is supported by Chinese Universities Scientific Fund Grant No. 2452018158. We would like to thank Dr. Wei He for helpful discussions.


\end{document}